\documentclass[secnumarabic,amssymb, amsmath, nofootinbib,tightenlines,
nobibnotes, aps, prl]{revtex4}
\usepackage{graphicx}
\begin{document}

\bigskip
\title{Annihilation contribution and $B\to a_0 \pi, f_0 K$ decays.}
\vskip 6ex
\author{D. Delepine}
\email{delepine@fisica.ugto.mx}
\author{J.\ L.\ Lucio M.}
\email{lucio@fisica.ugto.mx}
\affiliation{Instituto de F\'{\i}sica, Universidad de Guanajuato \\
Loma del Bosque \# 103, Lomas del Campestre, \\
37150 Le\'on, Guanajuato; M\'exico}
\author{Carlos A. Ram\'{\i}rez}
\email{mpramir@netscape.net}
\affiliation{Depto. de F\'{\i}sica, Universidad Industrial de Santander, \\
A.A. 678, Bucaramanga, Colombia}

\bigskip

\bigskip

\begin{abstract}
We analyze the decays $B^0 \to a^\pm_0 \pi^\mp$ and $B^{-,0} \to
f_0 K^{-,0}$ and show that within the factorization approximation
a phenomenological consistent picture can be obtained. We show
that in this approach the $O_6$ operator provides the dominant
contributions to the suppressed channel $B^0 \to a^+_0 \pi^-$.
When the $a_0$ is considered a two quark state, evaluation of the
annihilation form factor using Perturbative $QCD$ implies that
this contribution is not negligible, and furthermore it can
interfere constructively or destructively with other penguin
contributions. As a consequence of this ambiguity, the positive
identification of $B^0 \to \pi^+ a_0^-$ can not distinguish
between the two or four quark assignment of the $a_0$
\cite{suzuki}. According to our calculation, a best candidate to
distinguish the nature of $a_0$ scalar is $Br(B^-\to \pi^0a_0^-)$
since the predictions for a four quark model is one order of
magnitude smaller than for the two quark assignment. When the
scalars are seen as two quarks states, simple theoretical
assumptions based on $SU(2)$ isospin symmetry provide relations
between different $B$ decays involving one scalar and one
pseudoscalar meson.

\end{abstract}

\maketitle
\bigskip

\bigskip

\section{Introduction}

\bigskip

$B$ factories provides large samples of $B - \bar{B}$ mesons
allowing the study of physical phenomena such as CP violation, the
determination of the CKM mixing angles and the search for new
physics\cite{Bfactories,4}. Clearly, hadronic physics will
beneficiate of the high statistics achieved, and the study of
processes with small branching ratios will be possible. The full
understanding of the $B$ physics is still lacking as well as a
systematic first principles description of the phenomena involved.
Instead different theoretical approaches are compared to data and
assumptions such as factorization, or estimation of the relative
size of different contributions (tree level, annihilation,
penguins, final state interactions) can be tested. This can be
achieved in processes where the dominant contributions are
suppressed by symmetry or accidental cancellations.

\bigskip

The BABAR and Belle collaborations already reported precise
measurements of non-leptonic $B$ meson decays involving scalar
mesons with branching ratio of order as low as $10^{-6}$. Thus for
example, for the $B^0 \to f_0 K^0$ channel, besides the branching
ratio the CP violating asymmetries are reported and, from the two
pion spectrum, the authors are able to obtain the mass and width of
the $f_0 (980)$\cite{f0}. This is not the case for the $B\to a_0
(980)\pi$ where branching fraction for given final states are
reported -in particular $a^-_0 \pi^+ $- however in this case it is
not possible to separate the $B^0$ from the $\bar B^0$ decays,
unless a dominant decay mechanism is assumed\cite{ref12}. In this
context it is worth remarking that the $B\to a_0 (980)\pi$ decay was
suggested as a place where $\alpha$, the weak mixing angle, could be
measured through the CP violating asymmetries \cite{ref2}. However,
it was shown that the $B\to a^+_0 \pi^-$ is suppressed by $G$ parity
and also by isospin, which implies that in the symmetry limit no  CP
violating asymmetry is expected to be experimentally accessible
\cite{suzuki}. Thus, theoretical arguments support the idea that the
$B^0\to a^+_0 \pi^-$ is strongly suppressed, so that the reported
branching ratio can be identified with the dominant $B^0\to a^-_0
\pi^+$ decay. As a by product, in~ \cite{suzuki} the author
concludes that the positive identification of $B^0/\bar B^0\to
a^\pm_0 \pi^\mp$ is an evidence against the four quark assignment of
$a_0$ or else, for the breakdown of perturbative QCD. Both, the
suppression of the $B^0\to a^+_0 \pi^-$  channel as well as the
potential evidence in favor or against a four quark state are rather
interesting observation that deserves further analysis.
\bigskip

The low lying scalar sector of QCD represents a major challenge
\cite{ref4}. From the experimental point of view, the determination
of the nature of existing states has not been achieved while from
the theory side no consistent interpretation of the experimental
data exists \cite{ref5}. This is so even though a number of
processes involving the appropriated final state in the kinematical
region of interest have been analyzed. Thus, for example, $\phi \to
\pi\pi\gamma , J/\Psi \to \phi \pi\pi , \phi KK$ and central
production involve the $f_0 (980)$ and $a_0 (980)$ \cite{ref6},
whereas the di-pion in the $\Upsilon (nS) \to \Upsilon (mS)\pi\pi$
and $D\to \pi\pi\pi$ decays include the kinematical region where the
$f_0 (600)$ is expected to appear \cite{ref7,mass,oller}.
Unfortunately, although different processes are included in the
analysis, data is not good enough to provide a clear picture of the
scalars. In fact no consensus exist even on the fundamental
intrinsic properties (mass and width) of the low lying scalar mesons
\cite{4,mass}.
\bigskip

The appropriated theoretical description of non-leptonic $B$
decays involving scalar mesons is important not only to understand
the nature of the scalars but also because these must be
considered as background to other processes of interest in $B$
physics. Since scalars, vectors and tensors couple to two
pseudoscalars, the following decays lead to the same final state:
$B\to PV, B \to SP , B\to TP$ and $B\to PPP$, where $S,V,T$ and
$P$ stand for scalar, vector, spin 2 and pseudoscalar meson
respectively. B decays involving scalar mesons have been
considered by a number of authors. Thus for example in \cite{ref8}
the tree level Hamiltonian and quark model calculation are used to
predict branching ratios, while sum rules \cite{ref9} and
gluon-penguin dominance $(b\to sg)$ are the basis to interpret the
scalars produced in $B$ decays in terms of glue balls
\cite{ref10}, or still using QCD corrected Hamiltonian plus
factorization \cite{ref11} to propose evidence for two quark
content of the $f_0 (980)$.
\bigskip

In this note we analyze the $B^0 \to f_0 (980) K$ and $B\to
a_0(980) \pi$ decays using the factorization approximation. To
this end we use the $\Delta B=1$ weak Hamiltonian including QCD
corrections to next to leading order. To evaluate the matrix
elements we use values reported in the literature or model
dependent estimates of the decay constants and form factors. In
particular, the annihilation contribution is considered and
evaluated using Perturbative QCD. This is important in estimating
contributions previously neglected, and it is also relevant to
quantify the statement in \cite{suzuki} regarding the four quark
nature of the $a_0 (980)$.

\bigskip

\bigskip

\section{$\bar{B}^0 \to a^\pm_0 \pi^\mp$ and $B^{0,-} \to f_0 K^{0,-}$}

\bigskip

Following the conventional approach \cite{Buchala,Ali,chen,Keum,lu},
we start with the $\Delta B =1$ effective Hamiltonian $H_{eff}
(q=d,s)$
\begin{equation}
{\cal H}_{eff} = \frac{G_F}{\sqrt{2}} \left[ \lambda_{uq}(C_1 O^u_1 +C_2 O^u_2)
-\lambda_{tq} \left( \sum^{10}_{i=3} C_i O_i +C_g O_g\right) \right] +h.c.
\end{equation}

\noindent where $\lambda_{q^\prime q} = V_{q^\prime b} V^*_{q^\prime q},
 q=d,s, q^\prime = u,c,t, V_{ij}$ are the Cabibbo-Kobayashi-Maskawa (CKM)
matrix elements. The Wilson coefficients, $C_i$, including next to
leading order QCD corrections, are evaluated at the renormalization
scale $\mu\simeq m_B/2$. We use the conventions and $C_i$ values
reported in \cite{chen}. It remains to evaluate the matrix elements
between the states of interest.

\begin{equation}
A(B\to PS) =~< PS |H_{eff}|B>
\end{equation}

\noindent $P$ and $S$ stand for pseudoscalar and scalar meson respectively. In
terms of the amplitude the branching ratio is given by:
\begin{equation}
Br(B\to PS) \simeq \tau_B \frac{G^2_F|A(B\to PS)|^2}{32\pi m_B}
\end{equation}

\noindent with $\tau_B$ the appropriated B meson lifetime
$(\tau_{B^+} =1.65 \cdot 10^{-12} s, \tau_{B^0} = 1.56 \cdot
10^{-12} s)$.

\bigskip

The matrix elements are evaluated using the assumption of factorization.
In that approximation the matrix elements of interest are given by:

\begin{eqnarray}
A_{\bar B^0\to \pi^- a^+_0} &\simeq& \lambda_{ud} (a_1 X^{\pi
-}_{\bar B^0 a^+_0} + a_2 X^{\bar B^0}_{(a^+_0 \pi^-)_u})-
\lambda_{td} \left[ \left( a_4 + a_{10} - \frac{(a_6+a_8)m^2_\pi}{
\widehat{m}(m_b + m_u)}\right)
X^{\pi^-}_{\bar B^0 a^+_0} \right. \nonumber \\
& & \left. + \left( 2(a_3 - a_5) + a_4+ \frac{a_9-a_7-a_{10}}{2} -
\frac{(a_6 - a_8/2)m^2_B}{m_u (m_b + m_d)} \right) X^{\bar
B^0}_{(a^+_0 \pi^-)_u} \right]
\end{eqnarray}

\begin{eqnarray}
A_{\bar B^0\to \pi^- a^+_0} &\simeq& \lambda_{ud} (a_1
X^{a^-_0}_{\bar B^0 \pi^+} + a_2 X^{\bar B^0}_{(a^-_0 \pi^+)_u})- \lambda_{td}
\left[ ( a_4 + a_{10}) X^{a^-_0}_{\bar B^0 \pi^+} -2 (a_6+a_8)
\tilde X^{a^-_0}_{\bar B^0 \pi^+} \right. \nonumber \\
& & \left. + \left( 2(a_3 - a_5) + a_4+ \frac{a_9-a_7-a_{10}}{2} -
\frac{(a_6 - a_8/2)m^2_B}{m_u (m_d + m_b)} \right) X^{\bar
B^0}_{(a^-_0 \pi^+)_u} \right]
\end{eqnarray}

\begin{eqnarray}
A_{B^-\to \pi^0 a^-_0} &\simeq& \lambda_{ud} (a_1 (X^{a^-_0}_{B^-
\pi^0}+X^{B^-}_{a_0^-\pi^0}) + a_2 X^{ \pi^0_u}_{B^- a^-_0})-
\lambda_{td} \left[ ( a_4 + a_{10}) X^{a^-_0}_{ B^-
\pi^0} \right. \nonumber \\
& & -2 (a_6+a_8)\tilde X^{a^-_0}_{ B^- \pi^0}-
\left(a_4-\frac{3}{2}(a_9-a_7)-\frac{1}{2}a_{10}+\frac{(a_6+a_8)m_{\pi}^2}{m_u(m_b+m_d)}\right)
X^{\pi^0_u}_{B^-a_0^-} \nonumber \\
& &\left. +\left(
a_4+a_{10}+\frac{(a_6+a_8)m_B^2}{\widehat{m}(m_b+m_u)}X^{B^-}_{a_0^-\pi^0}\right)
\right]
\end{eqnarray}

\begin{eqnarray}
A_{B^-\to \pi^- a^0_0} &\simeq& \lambda_{ud} a_1 (X^{\pi^-}_{B^-
a^0_0}+X^{B^-}_{a_0^0\pi^-})- \lambda_{td} \left[ ( a_4 +
a_{10}-\frac{(a_6+a_8)m_{\pi}^2}{\widehat{m}(m_b+m_u)})X^{\pi^-}_{B^-a^0_0}
\right.
\nonumber \\
& & \left.
+\left(a_4+a_{10}-\frac{(a_6+a_8)m_B^2}{\widehat{m}(m_b+\widehat{m})}\right)
X^{B^-}_{a_0^0\pi^-} +(a_8-2a_6) \tilde{X}^{a_d^0}_{B^-\pi^-}
\right]
\end{eqnarray}

\begin{eqnarray}
A_{B^-\to f^0 K^-} &\simeq& \lambda_{us} a_1 \left[
X^{K^-}_{B^-f^0} + X^{B^-}_{f^0 K^-} \right] - \lambda_{ts}
\left[ \left( a_4 + a_{10}-\frac{2(a_6+a_8)m^2_B}{(m_b+m_u)(m_s +m_u)} \right)
X^{B^-}_{f^0 K^-} \right. \nonumber \\
& & \left. + \left(a_4 + a_{10} -\frac{2 (a_6 + a_8) m^2_K}{(m_u+
m_s) (m_b + m_u)} \right) X^{K^-}_{B^- f^0}
-(2a_6+a_8)\tilde{X}^{f_s^0}_{B^-K^-} \right]
\end{eqnarray}

\begin{eqnarray}
A_{\bar B^0\to f^0 \bar K^0} &\simeq& -\lambda_{ts} \left[ \left( (a_4
- \frac{a_{10}}{2} - \frac{(2a_6- a_8) m^2_B}{(m_b+ m_d)(m_s+ m_d)} \right)
X^{\bar B^0}_{f^0 \bar K^0} \right. \nonumber \\
& & \left. + \left( a_4 - \frac{a_{10}}{2} - \frac{(2a_6 -
a_8)m^2_K}{(m_s+m_d)(m_b+ m_d)} \right) X^{\bar K^0}_{\bar B^0 f^0}
 -(2a_6-a_8)\tilde{X}^{f_s^0}_{\bar B^0 \bar K^0}\right]
\end{eqnarray}

\noindent where $\widehat{m}=(m_u+m_d)/2$. For future reference in
Table 1 we quote the numerical values of the $a_i$ coefficients.
These expressions are obtained by inserting the vacuum between the
currents in all possible ways, and a typical $X^a_{b,c}$ product of
matrix elements is parameterized in terms of form factors and decay
constant as:
\begin{table}
\begin{tabular}{|c|c|c|c|c|c|c|c|c|c|c|}
  \hline
  References & $a_1$ & $a_2$ & $a_3$ & $a_4$ & $a_5$ & $a_6$ & $a_7$ &$a_8$ & $a_9$ & $a_{10}$ \\
  \hline
  \cite{Buchala} &  $1.039$& $0.084$ &  $40$& $-440$ & $-120$ & $-620$ & $0.7-i$ & $4.7-0.3 i$ & $-94-i$ & $-14-0.3i$ \\
  \cite{Ali} & $1.050$ & $0.053$ & $48$ & $-412-36i$& $-45$ & $-548-36i$ & $0.7-i$ & $4.7-0.3i$ & $-94-i$ & $-14-0.3i$ \\
  \cite{chen} & $1.046 $& $0.024$ & $72-0.3i$ & $-379-102i$ & $-27-0.3i$ & $-431-102i$ & $-0.81-2.4i$ & $3.3-0.8i$ & $-92.4-2.41i$ & $0.34-0.8i$ \\
  \cite{Keum} &$1.061$ &$0.011$&$63$     &$-317$ &$-60$     &$-473$ &$4$ &$5.4$ &$-87$     &$-2.4$  \\
  \cite{lu}&$1.029$ &$0.103$&$36$ &$-228$ &$-24$     &$-298$ & $12$     &$7.6$ &$-82$     &$-8.2$  \\
  \hline
\end{tabular}
\caption{Numerical values for the effective coefficientes
  $a_i^{eff}$ for $b \to d$ transitions at scale $\mu \approx m_b$ (for
  $a_3,...,a_{10}$ in units of $10^{-4}$ ($a_{2i-1}=C_{2i-1}+C_{2i}/N$, $
  a_{2i}=C_{2i}+C_{2i-1}/N$)) }
\end{table}
\begin{equation}
X^{\bar B^o}_{\bar K^0 f_0} = ~< \bar K^0 f^0 |(\bar
sd)_L|0><0|(\bar d b)_L| \bar B^0>~= f_B (m^2_{f^0} - m^2_K)
F^{f^0 \bar K^0}_0 (m^2_B)
\end{equation}
In the appendix we define in detail all of the $X^a_{b,c}$. Let us
start by summarizing our knowledge about the decay constants and
form factors entering the calculation. We can classify these in four
categories: Pseudoscalar decay constants $(f_\pi , f_K, f_B)$. The
values of the two former decay constants are taken from [4] while
for the later we use $f_B=170$ MeV\cite{fB}. The second kind are the
scalar decay constants $(f_a, f_f)$. For these we use published
values estimated using theoretical arguments \cite{fa}. We then have
form factors of the type $F^{SP}_0 (m^2_B)$, evaluated at the $m_B$
scale, {\it i.e.} calculable with perturbative methods. In Table
(\ref{table2}), we quote the values we use. Finally we need the form
factor $F^{BS}_0(m^2_P) , F^{BP}_0(m^2_S) $ where $S$ and $P$ stand
for a scalar or pseudoscalar meson. For the decay under
consideration these are evaluated at most at the scalar scale
(around 1 GeV), {\it i.e.} these can not be computed using
perturbative methods and, in general, few is known about their
values.
\begin{table}
\begin{tabular}{|c|c|c|}
  \hline
  $f_B$ & $170$ MeV & \cite{fB} \\
  $f_K$ & $159.8$ MeV & \cite{4} \\
  $f_{\pi}$ & $130.7$ MeV & \cite{4} \\
  $\tilde f_{f_s^0}$ & $180$ MeV  & \cite{fa} \\
  $\tilde f_{a_0}$ & $400$ MeV & \cite{fa} \\
  $F^{B^0\pi^-}_0$ & $0.28$ & \cite{FBpi} \\
$F^{B^0K^-}_0$ & $0.34$ & \cite{FBpi} \\

  \hline
\end{tabular}
\caption{ Numerical values of the form factors} \label{table2}
\end{table}

 For the $B\to f_0 K^0, B^- \to f_0 K^-$
and $B\to a_0 \pi$ we require $F^{B\pi}_0$, $F^{BK}_0$,
$F^{Bf^0}_0$, $F^{Ba_0}_0$, $F^{a\pi}_0$ and $F^{f^0K}_0$. The two
first ($F^{B\pi}_0$, $F^{BK}_0$) are relatively well-known and we
shall use the value presented in refs.\cite{FBpi}. In principle, the
other four form factors could be determined using experimental
data\footnote{ we used $Br(f^0 \to 2 \pi) =0.68$ \cite{4} in order
to get the $Br(B^{-,0} \to K^{-,0}f^0)$ from published results.}
\cite{f0,ref12,ref2} given in Table (\ref{table3}) but since the
experimental data involve large error bars, we prefer to evaluate
some of these form factors. In this paper we focus our interest on
the annihilation effects, for that reason we present an estimate of
$F^{a_0\pi}_0$ using perturbative QCD and considering $a_0$ as a two
or four quark state.

\begin{table}
\begin{tabular}{|c|c|c|}
  \hline
  $Br(\bar B^0 \to \pi^{\pm} a_0^{\mp})$ & $(2.8^{+1.65}_{1.47}) \ 10^{-6} $& \cite{ref12} \\
 $Br(\bar B^0-\to \pi^{-} a_0^{0})$ & $(3.6^{+2.25}_{-2.06}\ 10^{-6} $ & \cite{ref12} \\
  $Br(B^{-} \to K^{-}f^0)$ & $(13.5^{+3.6}_{-4.2})\ 10^{-6}$ & \cite{f0}\\
  $Br(\bar B^{0} \to K^{0}f^0)$ & $ (8.8 \pm 2.38)\ 10^{-6}$  & \cite{f0} \\
  \hline
\end{tabular}
\caption{ Branching ratios of measured PS channel decays of $B$
mesons} \label{table3}
\end{table}

\bigskip

\bigskip

\section{Annihilation form factors from Perturbative QCD.}

\bigskip

At tree level the $\bar B^0 \to \pi^+ a^-_0$ is strongly
suppressed due to the absence of second class currents. To get an
evaluation for such a decay an estimate of the contribution of the
$B$ annihilation is necessary. The annihilation amplitude is
proportional to $X^{\bar B^0}_{(a^-_0 \pi^+)_u}$ which itself is
proportional to the $F^{a^-_0 \pi^+}_0 (m^2_B)$ form factor. At
the scale $m^2_B$, perturbative QCD provides an adequate framework
to evaluate  this form factor. So, below we compute this form
factor using the standard approach of perturbative QCD \cite{PQCD}
assuming the scalar meson $a^-_0$ is a two quark state.

\bigskip

\bigskip


\bigskip

In order to fix the convention, we recall the form factor
definitions :

\begin{eqnarray}
\langle M_2 (p_2)|L^u|M_1(p_1)\rangle &=& \left( p_1+p_2 - \frac{m^2_1-m^2_2}{q^2} \right)_u F^{M_2M_1}_+ (q^2) + \left(\frac{m^2_1- m^2_2}{q^2}\right)
q_u F^{M_1 M_2}_0 (q^2), \\
\end{eqnarray}

\noindent with $q=p_1-p_2$. Projecting the amplitude on $q$ one
obtains:

\begin{eqnarray}
q_u \langle M_2 (p_2)|L^u|M_1 (p_1)\rangle &=& (m^2_1 - m^2_2)
F^{M_1, M_2}_0 (q^2) \\
q_u \langle M_2 (p_2)M_1 (p_1) |L^u|0 \rangle &=& (m^2_2 - m^2_1)
F^{M_2, M_1}_0 (q^2)
\end{eqnarray}

\noindent $PQCD$ contributions to both amplitudes have exactly the
same structure, so we compute $q_u \langle M_2 (p_2) |L^u| M_1
(p_1)\rangle$ following ref. \cite{du} then $q_u \langle M_2
(p_2)|L^u| M_1 (p_1) \rangle$ is obtained just changing the sign
of $p_2$. The form factors are expressed in terms of the
distribution amplitudes:

\begin{eqnarray}
\Psi_\pi (x,p) &=& \frac{-iI_c}{\sqrt{2N_c}} \phi_\pi (x) (\hat p +m_\pi )
\gamma_5 \\
\Psi_{a_0} (x,p) &=& \frac{I_c}{\sqrt{2N_c}} \phi_{a_0} (x) (\hat p +m_{a_0})
\gamma_5
\end{eqnarray}

\noindent where $I_C$ is the identity in color space, $\hat p
=~\gamma_\mu p^\mu$ and

\begin{eqnarray}
\int \phi_\pi (x) dx &=& \frac{1}{2\sqrt{2N_c}} f_\pi \\
\int \phi_{a_0} (x) dx &=& \frac{1}{2\sqrt{2N_c}} f_{a_0}
\end{eqnarray}

\noindent The wave functions $\phi_{\pi , a_0} (x)$ are given by
\cite{dielh}.

\begin{eqnarray}
\phi_\pi (x) &=& \frac{2N_c}{2\sqrt{2N_c}} f_\pi x (1-x)+ \cdots \\
\phi_{a_0} (x) &=& \frac{2N_c}{2\sqrt{2N_c}} f_{a_0} x (1-x) \left( 1+B_1
C^{3/2}_1 (2x-1) \right)+\cdots
\end{eqnarray}

\noindent where $f_\pi = 130 MeV, f_{a_0} = 1 MeV, |B_1 f_{a_0}|
\simeq 75 MeV$, and $C^{3/2}_1 (2x-1)$ is the Gegenbauer
polynomial. Thus, the matrix element is expressed as:

\begin{eqnarray}
q_\mu \langle \pi (p_2) |L^\mu | a_0 (p_1) \rangle &=& -C (R)
\frac{T_r (I_C)}{2N_c} g^2_s \int dxdy \phi_{a_0} (x) \phi_\pi (y) \nonumber \\
& & \left\{ \frac{Tr\left[\gamma_5 (\hat p_2+m_\pi
)\gamma_\nu\hat{P_{1l}}q_{\mu}
L^\mu (\hat p_1 +m_{a_0}) \gamma^\nu\right]}{k^2 P^2_{1l}}+\right. \nonumber \\
& & \left. \frac{Tr \left[ \gamma_5 (\hat p_2 + m_\pi ) q_\mu L^\mu
\hat{P_{2l}} \gamma^\nu (\hat p_1 + m_{a_0}) \gamma_\nu \right]}{k^2 P^2_{2l}} \right\}
\end{eqnarray}

\noindent where $C(R) = 4/3, (p_1 - p_2)^2 = q^2 = m^2_B$,
$k=-xp_1+(1-y)p_2$, $P_{1l}=k+yp_2$, $P_{2l}=-k+(1-x)p_1$ and

\begin{eqnarray}
 P^2_{1l} &=& x^2 m^2_{a_0} + m^2_\pi + x (m^2_B - m^2_{a_0} - m^2_\pi ), \\
P^2_{2l} &=& (1-y) m^2_B + ym^2_{a_0}-m^2_\pi y(1-y).
\end{eqnarray}

\bigskip \noindent Integrating numerically, one gets

\begin{equation}
\left| F^{a^-_0 \pi^+}_0 (m^2_B) \right| \approx 0.004
\end{equation}

It is important to notice that the CP-conserving phase of the
annihilation contributions is not fixed since we only known the
absolute value of $B_1$.

\bigskip
\begin{figure}
\includegraphics[width=10cm]{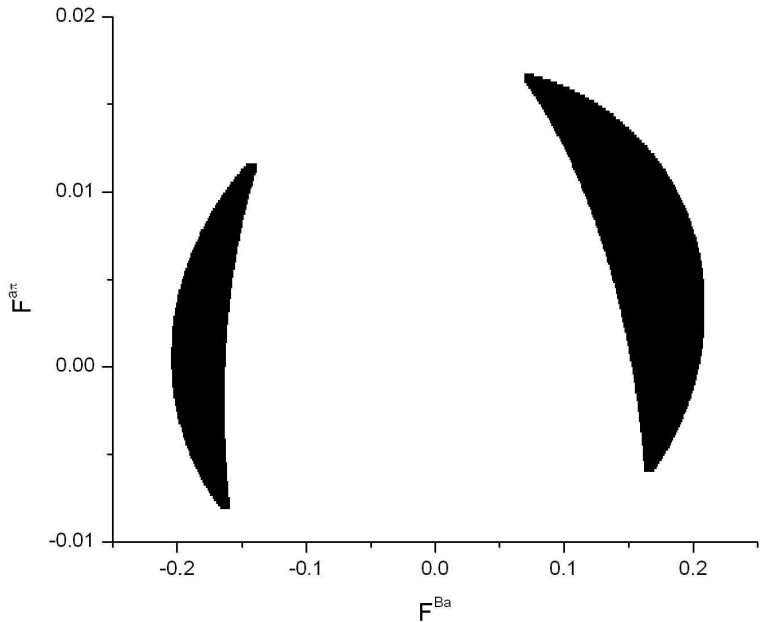}\\
  \caption{Allowed region for $F^{B^0a^+}_0$ and $F^{\pi a}_0$ at one $\sigma $
   using experimental data on $Br(B^0\rightarrow \pi^-a_0^+)$ and $Br(B^{-} \rightarrow \pi^-a_0^0)$}\label{fig2}
\end{figure}

\section{General framework to predict $B\to PS$}
One can proceed along similar lines for processes involving $a_0$
or $f_0$ scalar mesons. Instead, we use isospin, $SU(2)$ quark
symmetry and the quark contents of the scalar mesons to obtain
relation between the form factors. We also used available
experimental data to obtain constraints on the form factor values.
It turns out that the consistency of the two sets of values so
obtained provide further confidence on the approach.

\bigskip

We assume the conventional quark content of the pseudo scalar
mesons\cite{4} and parameterize the mixing in the scalar sector in
the strange-nonstrange basis as:





\bigskip

\begin{eqnarray}
\sigma &=& \cos \phi_S \bar n n - \sin \phi_S \bar s s , \\
f_0 &=& \sin \phi_S \bar n n + \cos \phi_S \bar s s ,
\end{eqnarray}

\noindent where $\bar n n = (\bar u u+ \bar d d)/\sqrt{2}$, and
the singlet-octet mixing angle $\theta_S$ is related to $\phi_S$
by  $\phi_S - \theta_S = \cos^{-1} [1/\sqrt{3}] \simeq 55^\circ$.
A diagrammatic analysis of the contributions of the form factors,
which involve the quark composition and isospin and $SU(2)$
symmetry between up and down quarks, lead the following  relations
:

\begin{equation}
\frac{\sqrt{2}}{\cos\phi_S} F^{B^- \sigma}_0 = \frac{\sqrt{2}}{\sin
\phi_S} F^{B^- f_0}_0 = \frac{\sqrt{2}}{\sin \phi_S} F^{\bar B^0
f_0}_0 = F^{\bar B^0 a^+_0}_0 \label{2q}
\end{equation}


\noindent {From} these relations it follows that $|F^{\bar B^0
f_0}_0|~ < ~$ $|F^{\bar B^0 a^+_0}_0|/\sqrt{2}$.

\bigskip
It is possible to obtain similar relations between the annihilation
form factors ($F^{a_0 \pi}_0$ and $F^{f^0K}_0$), however we will not
use $SU(3)$ symmetry since large deviations from the symmetry limit
are expected. Using the experimental results given in Table
(\ref{table3}), the values for the scalar masses given in
ref.\cite{4} and the values given in table (\ref{table2}) for the
other form factors appearing in the amplitudes, it is possible to
determine separately for each scalars $a_0$ and $f_0$ the allowed
regions for the values of the form factors $F^{a_0\pi}_0$
($F^{f_0K}_0$) and $F^{Ba_0}_0$ ($F^{BK}_0$) respectively.

\bigskip

\begin{figure}
\includegraphics[width=10cm]{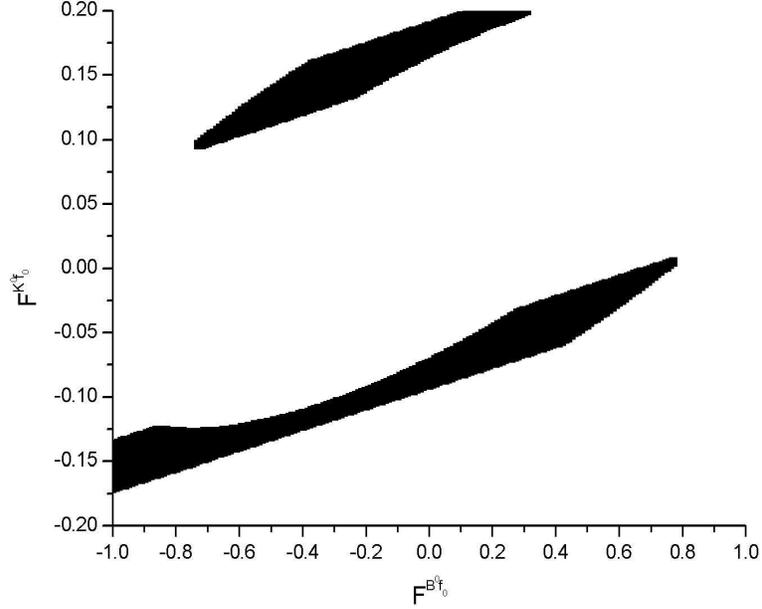}\\
  \caption{Allowed region for parameters $F^{B_0f_0}_0$ and $F^{K^0f_0}_0$ at one $\sigma $}\label{fig1}
\end{figure}

The results are summarized in figures \ref{fig2} and \ref{fig1}.
Assuming that Perturbative QCD leads
us the right order of magnitude for $F^{a_0\pi}_0$, it follows that:
\begin{equation}
0.14\leq|F^{Ba_0}_0|\leq 0.21
\end{equation}
We should note that this result is compatible (even if slightly
smaller) with the predictions for $F^{\bar B^0a_0^+}_0(0)=0.55\pm
0.22$ \cite{FBa} obtained in a model-dependent way. Using
Eq.(\ref{2q}), it follows that
\begin{equation}
|F^{\bar B^0 f_0}_0| \leq 0.20
\end{equation}
Figure (\ref{fig1}) shows the values allowed by the experimental
data when one standard  deviation is considered. One observes that
$|F^{\bar B^0 f_0}_0| \leq 0.20$ requires a large contribution from
$|F^{f_0K}_0|$. In fact, the smallest value for $|F^{f_0K}_0|$is
around $0.05$, which is more than one order of magnitude bigger than
the PQCD evaluation of $|F^{a_0\pi}_0|$. It is interesting to note
in this respect that:
\begin{equation}
\frac{|F^{f_0K}_0|}{|F^{a_0\pi}_0|} \approx \frac{m_K^2}{m_{\pi}^2}
\approx 12
\end{equation}

\subsection{Scalar mesons as $qq{\bar qq}$ states.}

Several models where the scalars are four quark states
\cite{maiani,mauro,black,oller} have been published but no model is
favored at present time. We shall apply our method to one example,
following \cite{maiani} we assume that the quarks contents of the
scalars is given by

\begin{eqnarray}
a_0^+ &=& uu\bar d\bar s,\hskip0.3cm a_0^-=ds\bar u\bar
s,\hskip0.3cm a_0^0={1\over \sqrt{2}}\left( us\bar
u\bar s-ds\bar d\bar s\right),\hskip0.3cm K_0^+=ud\bar d\bar s,\hskip0.3cm K_0^0=ud\bar u\bar s,\hskip0.3cm \bar K_0^0=us\bar u\bar d,\hskip0.3cm K_0^-=ds\bar u\bar d \nonumber \\
f_0 &=&{\cos \phi\over \sqrt{2}}\left(su\bar s\bar u+sd\bar s \bar
d\right)+\sin\phi\ ud \bar u \bar d, \hskip0.3cm \sigma=-{\sin
\phi\over \sqrt{2}}\left(su\bar s\bar u+sd\bar s \bar
d\right)+\cos\phi\ ud \bar u \bar d
\end{eqnarray}
where the mixing angle is obtained from the relation $\tan \phi
=-0.19$ (for $m_{\sigma}= 0.47$ GeV), so $\phi=-5.4^{\rm o}$ and
$84.6^{\rm o}$.
\bigskip

It is well-known that perturbative QCD predicts that the form factor
will go like $1/q^{2 (n-1)}$ where $n$ is the number of constituents
of the hadron. If $n=4$, $F^{a^-_0 \pi^+}_0 (m^2_B)$ is strongly
suppressed and annihilation can be neglected. Varying the
experimental results within one $\sigma$ and using the fact that in
four quark models for scalars the annihilation does not contribute
to the processes ($F^{a_0\pi}_0=F^{Kf_0}_0=0$), one concludes that:

\begin{equation}
0.70\leq F^{Bf_0}_0 \leq 0.75
\end{equation}

Proceeding in the same way with processes $B\to
a_0^0\pi^-,a_0^{\pm}\pi^{\mp}$, one gets
\begin{equation}
0.15 \leq|F^{Ba_0}_0|\leq 0.20
\end{equation}
which are closed to the values of $|F^{Ba_0}_0|$ obtained assuming
the scalars are two quark states.

\bigskip

\subsection{ Sub-dominant processes $Br(\bar B^0 \to \pi^+ a_0^-)$
and annihilation}

\begin{figure}
 \includegraphics[width=10cm]{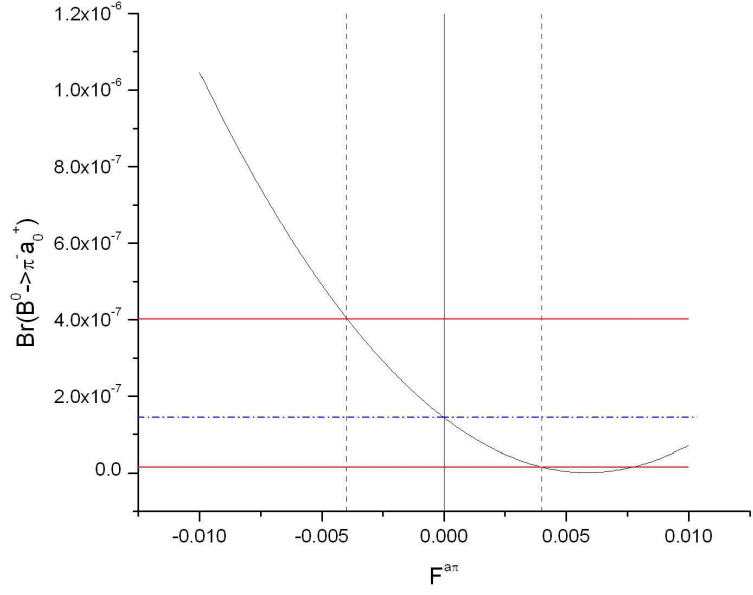}\\
  \caption{Branching ratio for $Br(\bar B^0 \to \pi^+ a_0^-)$. The dot-dashed line correspond to the four quark assignement for $a_0$, and
  the space between both solid horizontal line is the value for $Br(\bar B^0 \to \pi^+ a_0^-)$ expected from 2 quark models for $a_0$}\label{fig3}
\end{figure}
Once the allowed values for the form factors $F^{Bf_0}_0$ and
$F^{Ba_0}_0$ have been constrained, we turn our attention to the
subdominant processes $Br(\bar B^0 \to \pi^+ a_0^-)$ which is
strongly suppressed by G parity  and isospin. In \cite{suzuki} the
author concludes that a positive identification of this process is
an evidence against the four-quark assignment of $a_0$ or else for
breakdown of perturbative QCD. Using our estimates for annihilation
contributions obtained using PQCD one can predict the values of
$Br(\bar B^0 \to \pi^+ a_0^-)$. The results are presented in figure
\ref{fig3}. In the four quark models for $a_0$ where annihilation is
strongly suppressed, one gets for $Br(\bar B^0 \to \pi^+ a_0^-)
\approx 10^{-7}$. In the two quark model for $a_0$, the main source
of uncertainty is the phase of the annihilation contributions which
cannot be fixed using PQCD. Varying between $0$ and $\pi$ the
CP-conserving phase for the annihilation, it follows that:
\begin{equation}
10^{-9}\leq Br(\bar B^0 \to \pi^+ a_0^-)\leq 4\times 10^{-7}
\end{equation}
where the lower limit is obtained when annihilation interferes
destructively  and the upper limit when annihilation interferes
constructively with the other contributions. Our conclusion is
that this $B$ decay  cannot be used to distinguish between two and
four quark assignment of the $a_0$, unless one can obtain an
independent determination of the annihilation phase.

Another channel suppressed by G parity is $B^-\to \pi^0a_0^-$. It is
interesting that this channel is better to distinguish between the 2
or 4 quark models for $a_0$. Indeed, in 4 quark models for $a_0$,
with the value for $|F^{Ba_0}_0|$ determined in previous sections,
one gets
\begin{equation}
2\times 10^{-9} \leq Br(B^-\to \pi^0a_0^-) \leq 10^{-8}
\end{equation}

while in the 2 quark model:
\begin{equation}
6.4\times 10^{-8} \leq Br(B^-\to \pi^0a_0^-) \leq 2.4\times 10^{-7}
\end{equation}

\section{Conclusions}

In this paper we consider processes for which the dominant
contribution is suppressed.  Using the factorization approximation
and available experimental data, we estimate the effect of
annihilation penguins contribution to the processes $Br(\bar B^0
\to \pi^{\pm} a_0^{\mp})$ and $Br(\bar B^{0,-}\to K^{0,-} f_0)$.
We show that a consistent picture can be obtained when the scalars
are described as two quark states, although one requires an
important contributions from annihilation penguins to $Br(\bar
B^{0,-}\to K^{0,-} f_0)$. Within our analysis the four quark
models for $f_0$ cannot be excluded.
\bigskip

Applying our estimates of the annihilation contributions to
suppressed processes like  $Br(B^-\to \pi^0a_0^-)$ and $Br(\bar
B^0 \to \pi^+ a_0^-)$, we conclude that the positive
identification of $\bar B^0 \to \pi^+ a_0^-$ cannot be taken as
evidence for the four quark assignment of $a_0$. This is in
contrast with ref.\cite{suzuki}, where the annihilation
contribution is not quantified. Relevant for this conclusion is
the ambiguity in the CP-conserving phase of the annihilation
penguins contributions. Our best candidate process to distinguish
the nature of $a_0$ scalar is $Br(B^-\to \pi^0a_0^-)$ where the
predictions for 4 quark models are typically one order of
magnitude smaller than 2 quark models.
\bigskip

Using the mesons quark content, $SU(2)$ quark symmetry and isospin
we derive relations between the form factors $F^{Bf_0}_0$ to
$F^{Ba_0}_0$ for different charge states. One can extend the
analysis to $SU(3)$ however one expects large deviations from the
symmetry limit. This restricts the applicability of our approach
to the four quark states since in that kind of models scalars
necessarily involve strange quarks.

\bigskip


\section{Acknowledgements}

\bigskip

The work of D.D. was supported by project PROMEP/103.5/04/1335
(M\'exico) and the work of J.L.M.  was supported by CONACyT,
M\'exico under project 37234-E. C.A.R. wants to thank the hospitality of J.L.M. and D.D. at IFUG, where part of this work was done.

\bigskip

\section{appendix}

Below we list the  $X^a_{b,c}$ expressed in terms of the form
factors. We quote only those needed to compute the branching ratios
given in the paper:

\begin{eqnarray}
X^{\pi^-}_{\bar B^0 a_0^+} &=&<\pi^-|(\bar du)_L|0><a_0^+|(\bar ub)_L|\bar B^0>=f_\pi(m_B^2-m_a^2)F_0^{\bar B^0a_0^+}(m_\pi^2)\nonumber \\
X^{a_0^-}_{\bar B^0\pi^+} &= & <a_0^-|(\bar du)_L|0><\pi^+|(\bar u b)_L|\bar B^0>=-f_a(m_B^2-m_\pi^2)F_0^{\bar B^0\pi^+}(m_a^2)\nonumber \\
X^{a_0^-}_{B^-\pi^0} &=& <a_0^-|(\bar du)_L|0><\pi^0|(\bar u b)_L|B^->=f_a(m_B^2-m_\pi^2)F_0^{ B^-\pi^0}(m_a^2)\nonumber\\
X^{\pi^0_u}_{B^-a_0^-} &=& <\pi^0|(\bar uu)_L|0><a_0^-|(\bar d b)_L|B^->=\frac{f_{\pi}}{\sqrt{2}}(m_B^2-m_a^2)F_0^{B^-\pi^-}(m_\pi^2)\nonumber\\
X^{\pi^-}_{B^-S^0} &=&<\pi^-|(\bar ud)_L|0><S^0|(\bar ub)_L|B^->=f_\pi(m_B^2-m_{S^0}^2)F_0^{B^-S^0}(m_\pi^2)\nonumber \\
X^{B^-}_{S^0\pi^-} &= & <S^0P^-|(\bar du)_L|0><0|(\bar u b)_L|B^->=-f_B(m_{S_0}^2-m_\pi^2)F_0^{S_0\pi^- }(m_B^2)\nonumber \\
X^{B^-}_{a_0^-\pi^0} &=& <a_0^-\pi^0|(\bar du)_L|0><0|(\bar u b)_L|B^->=-f_B(m_a^2-m_\pi^2)F_0^{a_0^-\pi^0}(m_B^2)\nonumber \\
X^{\bar B^0}_{\bar K^0S_0} &=&<\bar K^0S^0|(\bar sd)_L|0><0|(\bar d b)_L|\bar B^0>=-f_B(m_{S^0}^2-m_K^2)F_0^{S^0\bar K^0}(m_B^2)\nonumber \\
X^{B^-}_{S_0K^-}& =& <S^0K^-|(\bar su)_L|0><0|(\bar u
b)_L|B^->=-f_B(m_S^2-m_K^2)F_0^{S^0K^-}(m_B^2)\nonumber \\
X^{\bar B^0}_{(a_0^-\pi^+)_u} &=& <a_0^-\pi^+|(\bar uu)_L|0><0|(\bar
d b)_L|\bar B^0>=-f_B(m_{a_0^-}^2-m_\pi^2)F_0^{a_0^-\pi^+}(m_B^2)\nonumber \\
X^{\bar B^0}_{a_0^+\pi^-} &=& <a_0^+\pi^-|(\bar uu)_L|0><0|(\bar d b)_L|\bar B^0>=-f_B(m_{a_0^-}^2-m_\pi^2)F_0^{a_0^+\pi^-}(m_B^2)\nonumber \\
X^{\bar K^0}_{\bar B^0S^0} &=&<\bar K^0|(\bar sd)_L|0><S^0|(\bar
db)_L|\bar B^0>=f_{\bar K^0}(m_B^2-m_{S^0}^2)F_0^{\bar
B^0S^0}(m_K^2)\nonumber \\
X^{K^-}_{B^-S^0} &=&<K^-|(\bar us)_L|0><S^0|(\bar
ub)_L|B^->=f_K(m_B^2-m_{S^0}^2)F_0^{B^-S^0}(m_\pi^2)\nonumber \\
\tilde X^{a_0^-}_{\bar B^0\pi^+}&=& <a_0^-|\bar du|0><\pi^+|\bar u
b|\bar B^0>=m_a\tilde f_{a_0^-}{m_B^2-m_\pi^2\over m_b-m_u}F_0^{\bar
B^0\pi^+}(m_a^2) \nonumber \\
\tilde X^{S_d^0}_{B^-\pi^-} &= & <S^0|\bar dd|0><\pi^-|\bar d
b|B^->={m_{S_0}\tilde{f}_{S_d^0}\over
m_b-m_d}(m_B^2-m_\pi^2)F_0^{B^-\pi^-}(m_{S_0}^2) \nonumber \\
\tilde X^{a_0^-}_{B^-\pi^0}&=&<a_0^-|\bar du|0><\pi^0|\bar u
b|B^->=m_a\tilde f_{a_0^-}{m_B^2-m_\pi^2\over m_b-m_d}F_0^{
B^-\pi^0}(m_a^2) \nonumber \\
\tilde X^{S_s^0}_{B^-K^-} &=& <S^0|\bar ss|0><K^-|\bar s
b|B^->=m_{S^0}\tilde f_{S^0_s}{m_B^2-m_K^2\over
m_b-m_s}F_0^{B^-K^-}(m_{S^0}^2)\nonumber \\
\tilde X^{S_s^0}_{\bar B^0\bar K^0} &=& <S^0|\bar ss|0><\bar
K^0|\bar s b|\bar B^0>=m_{S^0}\tilde f_{S^0_s}{m_B^2-m_K^2\over
m_b-m_s}F_0^{\bar B^0\bar K^0}(m_{S^0}^2)=\tilde X^{S_s^0}_{B^-K^-}
\end{eqnarray}
where $S_0$ is a neutral scalar ($a_0^0$ or $f^0$).

\section{Form factors definitions and conventions.}

\bigskip

In order to compute the amplitude using the factorization, we use
the following parametrization of the form factors. The decay
constants are defined as:

\begin{eqnarray}
\langle 0|A_\mu|P(q)\rangle &=& i f_P q_\mu \\
\langle 0|\bar q_1\gamma_5 q_2|P (q)\rangle &\simeq&
\frac{-if_Pm^2_P}{m_1+m_2}
\equiv \bar f_P m_P \\
\langle a^-_0 |\bar d \gamma_\mu u|0 \rangle &=& f_{a0} p_\mu \\
\langle a^-_0 |\bar d u|0 \rangle &=& m_{a0} \bar f_{a0}
\end{eqnarray}

Using the equations of motion $(-i \partial^\mu (\bar q_1 \gamma_\mu
\gamma_5 q_2) = (m_1+m_2) \bar q_1 \gamma_5 q_2$ and $-i\partial^\mu
(\bar q_1 \gamma_\mu q_2) = (m_1 - m_2) \bar q_1 q_2$\cite{Ali,chen}
on can show that $\bar f_S = m_S f_S/(m_1-m_2)$ and that $f_{S^0}
=0$ for a neutral scalar.

Form factors are defined as:

\begin{eqnarray}
<M_2 (p_2) |L_\mu |M_1 (p_1)> &=& \left( p_1 + p_2 -
\frac{m^2_1-m^2_2}{q^2}q \right)_\mu F^{M_1 M_2}_+ +
\frac{m^2_1-m^2_2}{q^2} q_\mu
F^{M_1M_2}_0 (q^2) \\
<M_2 (p_2) M_1 (p_1) |L_\mu |0> &=& \left( p_2 - p_1 -
\frac{m^2_2-m^2_1}{q^2}q \right)_\mu F^{M_2 M_1}_+ (q^2) +
\frac{m^2_2-m^2_1}{q^2} q_\mu F^{M_2M_1}_0 (q^2)
\end{eqnarray}

\noindent where $L_\mu = \gamma^\mu \frac{1-\gamma_5}{2} =
\gamma^\mu P_L$. A factor of $-i$ has to be added to the form
factors in the case one of the mesons is scalar.


\bigskip

\begin{references}
\bibitem{Bfactories} P.~Ball et al., hep-ph/0003238,
http://lhcb.web.cern.ch/lhcb; K. Anikeev et al., hep-ph/0201071,
http://www-btev.fnal.gov; http//www.hera-b.desy.de; A.G. Akeroyd
et al., hep-ex/0406071; I.I. Bigi and A.I. Sanda, hep-ph/0401003.
\bibitem{4} S. Eidelman {\it et al.} [Particle Data Group Coll.],
Phys. Lett. {\bf B592}, 1 (2004) pdg.lbl.gov.

\bibitem{f0}
B.~Aubert {\it et al.}  [BABAR Collaboration],
arXiv:hep-ex/0408095;
B.~Aubert {\it et al.}  [BABAR Collaboration],
arXiv:hep-ex/0406040;
A.~Garmash {\it et al.}  [Belle Collaboration],
Phys.\ Rev.\ D {\bf 69}, 012001 (2004)
[arXiv:hep-ex/0307082];
S.~Laplace  [BABAR Collaboration],
arXiv:hep-ex/0309049;
K.~Abe {\it et al.}  [Belle Coll.],
Phys.\ Rev.\ D {\bf 65}, 092005 (2002) [arXiv:hep-ex/0201007].

\bibitem{ref12}
B.~Aubert {\it et al.}  [BABAR Collaboration],
arXiv:hep-ex/0407013;
J.~G.~Smith,
arXiv:hep-ex/0406063;
B.~Aubert {\it et al.}  [BABAR Coll.],
arXiv:hep-ex/0107075.



\bibitem{ref2} S.~Laplace and V. Shelkov, Eur.Phys.J.C 22, 431
(2001) (hep-ph/0105252).
\bibitem{suzuki} M. Suzuki, Phys. Rev. D 65, 097501 (2002).
\bibitem{fa} K. Maltman, Phys. Lett. {\bf B462}, 14 (1999) [hep-ph/9906267];
Y.~Meurice,
Phys.\ Rev.\ D {\bf 36}, 2780 (1987);
Mod.\ Phys.\ Lett.\ A {\bf 2}, 699 (1987);
C.~M.~Shakin and H.~Wang,
Phys.\ Rev.\ D {\bf 63}, 074017 (2001);
F.~De Fazio and M.~R.~Pennington,
Phys.\ Lett.\ B {\bf 521}, 15 (2001)
[arXiv:hep-ph/0104289].

\bibitem{ref4} see for example,
S. Spanier and N.A. Tornqvist, Phys. Lett. B 592:1 (2004), C.
Amsler and N.A. Tornqvist Phys. Rep. 389, 61 (2004) and references
therein.

\bibitem{ref5} F.~E.~Close and N.~A.~Tornqvist,
J.\ Phys.\ G {\bf 28}, R249 (2002)[arXiv:hep-ph/0204205];
A.~Abdel-Rehim, D.~Black, A.~H.~Fariborz and J.~Schechter,
AIP Conf.\ Proc.\  {\bf 687}, 51 (2003); E. van Beveren and G. Rupp
Mod. Phys. Lett. {\bf A19}, 1949 (2004); M. Uehara hep-ph/0404221;
M.~R.~Pennington,
eConf {\bf C030614}, 002 (2003)
[arXiv:hep-ph/0310186];
J.~L.~Lucio Martinez and M.~Napsuciale,
Phys.\ Lett.\ B {\bf 454} (1999) 365 [arXiv:hep-ph/9903234];
J.~L.~Lucio Martinez and J.~Pestieau,
Phys.\ Rev.\ D {\bf 42} (1990) 3253;
N.~N.~Achasov and V.N. Ivanchenko, Nucl. Phys. {\bf B315} (1989)465;
F.E. Close, N. Isgur, S. Kumano, Nucl. Phys. {\bf B389}, 513 (1993);
V.~V.~Anisovich and A.~V.~Sarantsev, Eur.\ Phys.\ J.\ A {\bf 16},
229 (2003) [arXiv:hep-ph/0204328];
F.~De Fazio and M.~R.~Pennington,
Phys.\ Lett.\ B {\bf 521} (2001) 15 [arXiv:hep-ph/0104289].

\bibitem{ref6}
R.~Akhmetshin {\it et al.}  [CMD-2 Collaboration],
Phys.\ Lett.\ B {\bf 462} (1999) 380 [arXiv:hep-ex/9907006].
V.~M.~Aulchenko {\it et al.}  [SND Collaboration],
Phys.\ Lett.\ B {\bf 440} (1998) 442 [arXiv:hep-ex/9807016].
M.~N.~Achasov {\it et al.},
Phys.\ Lett.\ B {\bf 485} (2000) 349 [arXiv:hep-ex/0005017].
J.~E.~Augustin {\it et al.}  [DM2 Collaboration],
Nucl.\ Phys.\ B {\bf 320} (1989) 1.


\bibitem{ref7}
E.~M.~Aitala {\it et al.}  [E791 Collaboration],
Phys.\ Rev.\ Lett.\  {\bf 89} (2002) 121801 [arXiv:hep-ex/0204018];
M.~Ishida, S.~Ishida, T.~Komada and S.~I.~Matsumoto,
Phys.\ Lett.\ B {\bf 518} (2001) 47;
H.~Muramatsu {\it et al.}  [CLEO Collaboration],
Phys.\ Rev.\ Lett.\  {\bf 89} (2002) 251802 [Erratum-ibid.\  {\bf
90} (2003) 059901] [arXiv:hep-ex/0207067].


\bibitem{mass}
D.~V.~Bugg,
Phys.\ Lett.\ B {\bf 572} (2003) 1 [Erratum-ibid.\ B {\bf 595}
(2004) 556];
G.~Colangelo, J.~Gasser and H.~Leutwyler,
Nucl.\ Phys.\ B {\bf 603} (2001) 125 [arXiv:hep-ph/0103088];
R.~Escribano, A.~Gallegos, J.~L.~Lucio M, G.~Moreno and J.~Pestieau,
Eur.\ Phys.\ J.\ C {\bf 28} (2003) 107 [arXiv:hep-ph/0204338],
A.~Gallegos, J.~L.~Lucio M. and J.~Pestieau,
Phys.\ Rev.\ D {\bf 69} (2004) 074033 [arXiv:hep-ph/0311133].

\bibitem{oller}
J.~A.~Oller, E.~Oset and J.~R.~Pelaez,
Phys.\ Rev.\ D {\bf 59} (1999) 074001 [Erratum-ibid.\ D {\bf 60}
(1999) 099906] [arXiv:hep-ph/9804209];
J.~A.~Oller,
Nucl.\ Phys.\ A {\bf 727} (2003) 353 [arXiv:hep-ph/0306031].

\bibitem{ref8}
 V.~Chernyak,
Phys.\ Lett.\ B {\bf 509}, 273 (2001) [arXiv:hep-ph/0102217].\\
A.~C.~Katoch and R.~C.~Verma,
Int.\ J.\ Mod.\ Phys.\ A {\bf 11}, 129 (1996).

\bibitem{ref9}
see for examples, T.~V.~Brito, F.~S.~Navarra, M.~Nielsen and
M.~E.~Bracco,
arXiv:hep-ph/0411233;
A.~Gokalp, Y.~Sarac and O.~Yilmaz,
arXiv:hep-ph/0410380;
H.~G.~Dosch, E.~M.~Ferreira, F.~S.~Navarra and M.~Nielsen,
Phys.\ Rev.\ D {\bf 65} (2002) 114002 [arXiv:hep-ph/0203225].
A.~Ali, V.~M.~Braun and H.~Simma,
Z.\ Phys.\ C {\bf 63} (1994) 437 [arXiv:hep-ph/9401277].
V.~M.~Braun and A.~V.~Kolesnichenko,
Phys.\ Lett.\ B {\bf 175} (1986) 485 [Sov.\ J.\ Nucl.\ Phys.\  {\bf
44} (1986\ YAFIA,44,756-773.1986) 489.1986\ YAFIA,44,756].


\bibitem{ref10}
P.~Minkowski and W.~Ochs,
arXiv:hep-ph/0404194;
arXiv:hep-ph/0304144.;
Nucl.\ Phys.\ Proc.\ Suppl.\  {\bf 121}, 123 (2003)
[arXiv:hep-ph/0209225];
Nucl.\ Phys.\ Proc.\ Suppl.\  {\bf 121}, 119 (2003)
[arXiv:hep-ph/0209223];
Eur.\ Phys.\ J.\ C {\bf 9}, 283 (1999) [arXiv:hep-ph/9811518].\\
E.~Klempt, B.~C.~Metsch, C.~R.~Munz and H.~R.~Petry,
Phys.\ Lett.\ B {\bf 361}, 160 (1995) [arXiv:hep-ph/9507449].\\
E.~Klempt,
arXiv:hep-ex/0101031.\\
V.~V.~Anisovich and A.~V.~Sarantsev,
Eur.\ Phys.\ J.\ A {\bf 16}, 229 (2003) [arXiv:hep-ph/0204328].

\bibitem{ref11}
C.~H.~Chen,
Phys.\ Rev.\ D {\bf 67}, 094011 (2003) [arXiv:hep-ph/0302059];
Phys.\ Rev.\ D {\bf 67}, 014012 (2003) [arXiv:hep-ph/0210028].
H.~Y.~Cheng,
Phys.\ Rev.\ D {\bf 67}, 034024 (2003)[arXiv:hep-ph/0212117].


\bibitem{Ali}A. Ali, G. Kramer and C.D. Lu, Phys. Rev. D {\bf 58}, 094009
           (1998) [arXiv:hep-ph/9804363];
           A. Ali and C. Greub, Phys. Rev. D {\bf 57}, 2996 (1998)
       [arXiv:hep-ph/9707251].
\bibitem{Buchala} M. Beneke, G. Buchalla, M. Neubert and C.T. Sachrajda, Nucl. Phys.
       B {\bf 606}, 245 (2001) [arXiv:hep-ph/0104110];
       G. Buchalla, A.J. Buras and M.E. Lautenbacher, Rev. Mod. Phys.
       {\bf 68}, 1125 (1996) [arXiv:hep-ph/9512380].
\bibitem{chen}Y.H. Chen, H.Y. Cheng, B. Tseng and K. C. Yang, Phys. Rev. D
       {\bf 60}, 094014 (1999) [arXiv:hep-ph/9903453].
\bibitem{Keum}Y.Y. Keum and H. n. Li, Phys. Rev. D {\bf 63}, 074006 (2001)
       [arXiv:hep-ph/0006001];
       Y.Y. Keum, H.N. Li and A.I. Sanda, Phys. Rev. D {\bf 63}, 054008
       (2001) [arXiv:hep-ph/0004173];
       Y.Y. Keum and A.I. Sanda, Phys. Rev. D {\bf 67}, 054009 (2003)
       [arXiv:hep-ph/0209014];
       Y.Y. Keum, H. n. Li and A.I. Sanda, Phys. Lett. B {\bf 504}, 6
       (2001) [arXiv:hep-ph/0004004].

\bibitem{lu}C.D. Lu, K. Ukai and M.Z. Yang, Phys. Rev. D {\bf 63}, 074009
       (2001) [arXiv:hep-ph/0004213];
       C.D. Lu and M.Z. Yang, Eur. Phys. J. C {\bf 23}, 275 (2002)
       [arXiv:hep-ph/0011238];
       C.D. Lu and D.X. Zhang, Phys. Lett. B {\bf 400}, 188 (1997)
       [arXiv:hep-ph/9703439].



\bibitem{fB} M.~Guagnelli, F.~Palombi, R.~Petronzio and N.~Tantalo,
Nucl.\ Phys.\ Proc.\ Suppl.\  {\bf 119}, 616 (2003)
[arXiv:hep-lat/0209113];
D.~Becirevic,
arXiv:hep-ph/0310072.









\bibitem{FBpi} A.~Abada, D.~Becirevic, P.~Boucaud, J.~P.~Leroy, V.~Lubicz and
F.~Mescia, 
mesons from lattice  QCD,'' Nucl.\ Phys.\ B {\bf 619}, 565 (2001)
[arXiv:hep-lat/0011065];
D.~Becirevic, S.~Prelovsek and J.~Zupan,
Phys.\ Rev.\ D {\bf 68}, 074003 (2003) [arXiv:hep-lat/0305001];
D. Be\'crevi{\'c}, S. Prelov$\check{\rm s}$ek and Ju. Zupan, Phys.
Rev. {\bf D67}, 054010 (2003); ICHEP-02, Amsterdam 2002;
P.~Ball and R.~Zwicky,
arXiv:hep-ph/0406261



\bibitem{PQCD} G.~P.~Lepage and S.~J.~Brodsky,
Phys.\ Rev.\ D {\bf 22}, 2157 (1980);
 G.~R.~Farrar and D.~R.~Jackson,
Phys.\ Rev.\ Lett.\  {\bf 43}, 246 (1979);
S. Brodsky, Lett. Nuov. Cim. {\bf 7}, 719 (1973);
S.~J.~Brodsky and G.~R.~Farrar,
Phys.\ Rev.\ D {\bf 11}, 1309 (1975);
Phys.\ Rev.\ Lett.\  {\bf 31}, 1153 (1973).


\bibitem{du}D.~s.~Du, D.~Yang and G.~Zhu,
Phys.\ Rev.\ D {\bf 60}, 054015 (1999).


\bibitem{dielh}M. Diehl and G. Hiller, JHEP {\bf 0106}, 067 (2001) [arXiv:hep-ph/0105194].

\bibitem{FBa} A.~Deandrea and A.~D.~Polosa, Phys.\ Rev.\ Lett.\  {\bf 86}, 216 (2001)
[arXiv:hep-ph/0008084]. \\ N.~Paver and Riazuddin,
arXiv:hep-ph/0107330.


\bibitem{maiani} L.~Maiani, F.~Piccinini, A.~D.~Polosa and V.~Riquer,
arXiv:hep-ph/0407017.





\bibitem{black}
D.~Black, A.~H.~Fariborz, F.~Sannino and J.~Schechter,
Phys.\ Rev.\ D {\bf 59} (1999) 074026 [arXiv:hep-ph/9808415].



\bibitem{mauro}
M.~Napsuciale and S.~Rodriguez,
Int.\ J.\ Mod.\ Phys.\ A {\bf 16} (2001) 3011
[arXiv:hep-ph/0204149].













\end{references}
\end{document}